\documentclass[%
 aip,
 apl,
 amsmath,amssymb,
 reprint,%
]{revtex4-1}
\usepackage{comment}
\usepackage{graphicx}
\usepackage{dcolumn}
\usepackage{bm}
\usepackage{hyperref}
\usepackage{physics}
\usepackage{standalone}
\usepackage{ulem}
\usepackage{import}

\usepackage[utf8]{inputenc}
\usepackage[T1]{fontenc}
\usepackage{mathptmx}
\usepackage{etoolbox}

\usepackage{xr}
\makeatletter

\newcommand*{\addFileDependency}[1]{
\typeout{(#1)}
\@addtofilelist{#1}
\IfFileExists{#1}{}{\typeout{No file #1.}}
}\makeatother

\newcommand{\orcid}[1]{\href{https://orcid.org/#1}{\textcolor[HTML]{A6CE39}{\aiOrcid}}}

\usepackage{chngcntr}

\usepackage{tikz,xcolor}

\definecolor{lime}{HTML}{A6CE39}
\DeclareRobustCommand{\orcidicon}{%
	\begin{tikzpicture}
	\draw[lime, fill=lime] (0,0) 
	circle [radius=0.16] 
	node[white] {{\fontfamily{qag}\selectfont \tiny ID}};
	\draw[white, fill=white] (-0.0625,0.095) 
	circle [radius=0.007];
	\end{tikzpicture}
	\hspace{-2mm}
}

\foreach \x in {A, ..., Z}{%
	\expandafter\xdef\csname orcid\x\endcsname{\noexpand\href{https://orcid.org/\csname orcidauthor\x\endcsname}{\noexpand\orcidicon}}
}

\newcommand{\modify}[1]{\textcolor{black}{#1}}
\newcommand{\modifysecond}[1]{\textcolor{black}{#1}}

\makeatletter
\def\@email#1#2{%
 \endgroup
 \patchcmd{\titleblock@produce}
  {\frontmatter@RRAPformat}
  {\frontmatter@RRAPformat{\produce@RRAP{*#1\href{mailto:#2}{#2}}}\frontmatter@RRAPformat}
  {}{}
}%
\makeatother

\begin{document}

\preprint{APS/123-QED}

\title{Investigation of Reflection-Based Measurements of Microwave Kinetic Inductance Detectors in the Optical Bands}
\author{Jie Hu\orcidA{}}
\affiliation{GEPI, Observatoire de Paris, PSL Université, CNRS, 75014 Paris, France} 
\affiliation{Université de Paris, CNRS, Astroparticule et Cosmologie, F-75013 Paris, France}

\author{Faouzi Boussaha\orcidD{}}
\affiliation{GEPI, Observatoire de Paris, PSL Université, CNRS, 75014 Paris, France} 
\author{Paul Nicaise\orcidC{}}
\affiliation{GEPI, Observatoire de Paris, PSL Université, CNRS, 75014 Paris, France}

\author{Christine Chaumont\orcidG{}}
\affiliation{GEPI, Observatoire de Paris, PSL Université, CNRS, 75014 Paris, France}

\author{Maria Appavou}
\affiliation{GEPI, Observatoire de Paris, PSL Université, CNRS, 75014 Paris, France}

\author{Viet Dung Pham\orcidF{}}
\affiliation{Université de Paris, CNRS, Astroparticule et Cosmologie, F-75013 Paris, France}

\author{Michel Piat\orcidH{}}
\affiliation{Université de Paris, CNRS, Astroparticule et Cosmologie, F-75013 Paris, France}

\email[Authors to whom correspondence should be addressed:]{jie.hu@obspm.fr}

\date{\today}

\begin{abstract}
In this paper, we investigate the single photon response from the reflection of the Microwave Kinetic Inductance Detector (MKID) array. 
Reflection measurements are carried out using two configurations: one is measured simultaneously with the transmission, and the other is obtained with a single-ended MKID array terminated with an open load. Compared with the transmission, reflection measurements significantly reduce the readout noise of the single-ended MKID array. This is also reflected in the improvement of the median energy resolving power by around 20-30\% under pulsed photon illumination at $\lambda = 405$~nm, mainly due to an increase in the size of the resonance circle on the IQ plane. This method has the potential to be used to read out large MKID arrays. 
\end{abstract}

\maketitle

The Microwave Kinetic Inductance Detector\cite{Day2003} (MKID) has emerged as a groundbreaking technology with wide-ranging applications in astrophysics, spanning from millimeter-wave\cite{NIKA2018} and optical\cite{Meeker2018, Hu2023} to X-ray observations\cite{Swimmer2023} for its wideband operation, intrinsic frequency domain multiplexity and relatively simple fabrication. In the optical and near-infrared bands, MKIDs demonstrate a distinctive blend of excellent energy resolution and the ease of scalability into large arrays\cite{Mazin2012, Gao2012, visser2021, Zobrist2022}, providing a significant advantage in applications like extreme faint object detection and low-resolution spectroscopy like SPIAKID\cite{Hu2023}.

MKIDs typically consist of high-quality lumped superconducting resonators in the optical and near-infrared bands. These resonators, as discussed in \cite{Beldi2019, visser2021, HuJie2021, Zobrist2022, Nicaise2022}, are constructed with an interdigitated capacitor (IDC) and a meander inductor. When photons with energies $h\nu > 2\Delta$ (2$\Delta$ the Cooper pair binding energy) are absorbed in the meander, they generate phonons, disrupt Cooper pairs, and produce quasiparticles (QPs). These additional QPs increase the kinetic inductance, lowering the resonator's resonance frequency and quality factor. This response can be detected through changes in the phase and amplitude of the resonator, respectively, by applying a probing tone at the resonance frequency.

The performance of optical MKIDs is characterized by their energy-resolving power $R$, which currently faces domination by three distinct noise sources. The first is the noise caused by the escape of hot phonons from the meander to the substrate, as discussed in the literature\cite{visser2021, Zobrist2022}. The second is the two-level system noise (TLS) generated within the substrate and at the superconductor/substrate interface  \cite{Gao2008}. The last is the readout noise from the system, primarily stemming from the low noise amplifier (LNA), which plays a significant role. Efforts have been dedicated to reducing this readout noise, notably through implementing superconducting parametric amplifiers\cite{Zobrist2019, Zobrist2022}, which are still not commercially available \modify{for reading out MKIDs.}

\begin{figure}
    \centering
    \includegraphics[width = \columnwidth]{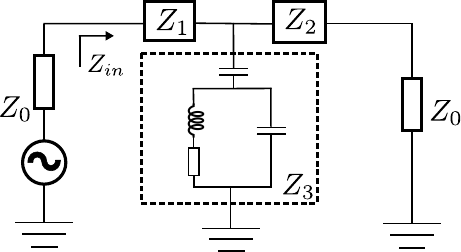}
    \caption{MKIDs T-junction circuit}
    \label{fig: T-junction circuit}
\end{figure}

Presently, MKIDs are typically read out via the transmission of the resonator. However, due to the inherent nature of MKIDs, strong reflections can occur at resonance, especially in the case of high-quality superconducting resonators where the reflected signal ($S_{11}$) can surpass the transmitted signal ($S_{21}$). Consequently, by utilizing $S_{11}$ for resonator readout, there is potential to diminish the sensitivity of the resonator to readout noise within the system. This signal, however, is not traditionally measured in the conventional homodyne configuration.
\begin{figure*}
    \centering
    \includegraphics[width = \textwidth] {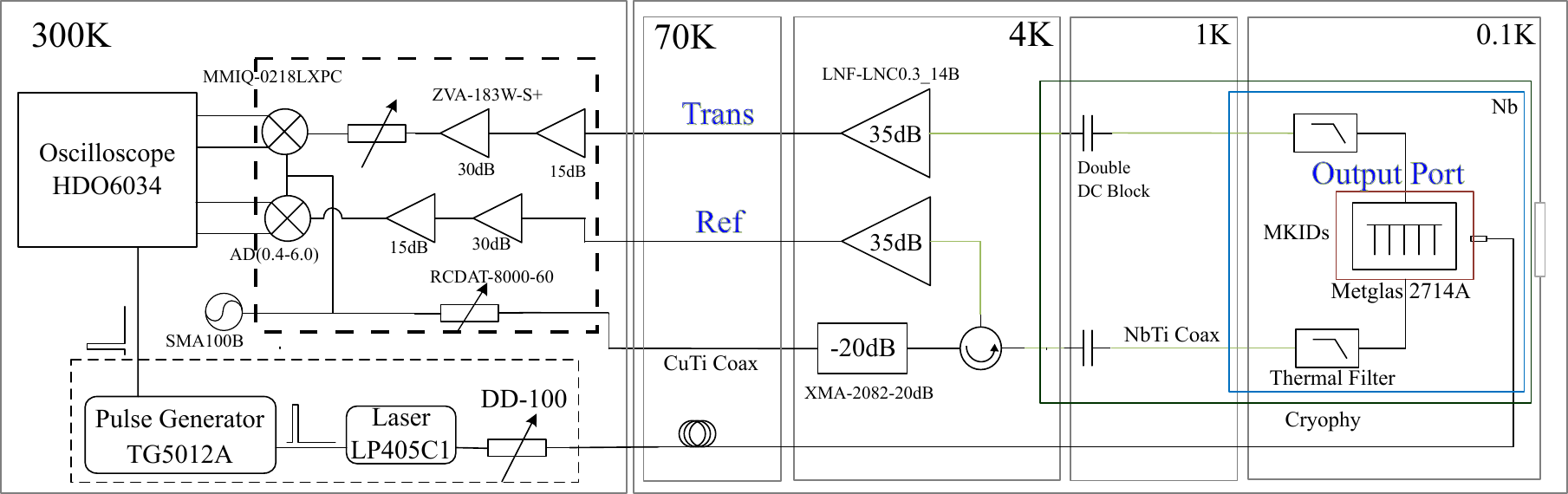}
    \caption{Measurement Setup with reflection ($S_{11}$) and transmission ($S_{21}$). The $S_{11}$-open is obtained by terminating the output port of the MKIDs with an open load. The reflected signal is obtained by putting an isolator at the input of the resonator. The isolator is currently placed on the 4K stage due to the limited space on the 0.1K stage.}
    \label{fig: Measurement Setup}
\end{figure*} 

In this paper, we investigate the single-photon response of TiN/Ti/TiN MKIDs\cite{Hu2023} measured by reflection. To capture the reflected signal, a cryogenic circulator is positioned at the input of the MKID array, with the reflected signal ($S_{11}$) read from the circulator's third port—a methodology commonly employed for qubit readout. \cite{Wang2014, Wang2021}. Two distinct configurations are compared. In the first configuration, both $S_{11}$ and $S_{21}$, taken as a reference, are measured simultaneously. In contrast, only $S_{11}$ is measured in the second configuration when the MKID array is terminated with an open load. The latter is referred to as single-ended MKIDs. Across all configurations, we observed a single-photon response, with the single-ended MKID array demonstrating the most favorable median $R$, approximately 20-30\% 

The equivalent circuit of the MKID shunted-coupled to a transmission line with a characteristic impedance $Z_0$ can be generally represented in Fig.~\ref{fig: T-junction circuit}. \modify{$Z_1$ and $Z_2$ are general descriptions of the port impedance, which can either be the impedance mismatch or the impedance of the transmission lines.} $Z_3$ is the MKID impedance.
$S_{11}$ can be obtained as 
\begin{align}\label{eqn: S11 transformed 1}
    S_{11} = \frac{A Z_3 + \Gamma_0}{1 + 2Z_3/\hat{Z}_0}
\end{align}
with 
\begin{align}
    \hat{Z}_0 &= \frac{2(Z_1 + Z_0)(Z_2 + Z_0)}{2Z_0 + Z_1 + Z_2} \\
    A &= \frac{Z_1 + Z_2}{(Z_0 + Z_1)(Z_0 + Z_2)} \\
    \Gamma_0 &= \frac{Z_1 - Z_0}{Z_1 + Z_0}
\end{align} 
$\hat{Z}_0$ and $\Gamma_0$ are the equivalent characteristic impedance of the transmission line and the reflection coefficient at the MKID array input. $S_{21}$ of the MKIDs can be expressed as\cite{Zobrist2022}
\begin{align}\label{eqn: S21}
    S_{21} = \frac{2Z_0}{2Z_0 + Z_1 + Z_2}(1 - \frac{1}{1 + 2Z_3/\hat{Z}_0})
\end{align}
with\cite{Khalil2012,Zobrist2022}
\begin{align}\label{eqn: 1-2Z_3}
    \frac{1}{1 + 2Z_3/\hat{Z_0}} = \frac{Q/Q_c - 2jQx_a}{1+2jQx_g}
\end{align}
where $Q$ and $Q_c$ are, respectively, the quality factor and the coupling quality factor of the resonator, $x_g = (f-f_r)/f_r$ is the detuning of the generator frequency from resonance, and $x_a = \delta f_a/f_r$ is the fractional detuning of the original resonance frequency due to the impedance mismatches. By substituting Eq.~(\ref{eqn: 1-2Z_3}) into Eq.~(\ref{eqn: S11 transformed 1}), 
\begin{align}
    S_{11} &= \gamma - (\gamma - \Gamma_0)\frac{Q/Q_c - 2jQx_a}{1+2jQx_g}
\end{align}
Then, $S_{11}$ can be fitted with 
\begin{align} \label{eqn: S11 final}
    S_{11} = (\gamma - \Gamma_0) ( \gamma_0 - \frac{Q/Q_c - 2jQx_a}{1+2jQx_g})   
\end{align}
with $\gamma_0 = \gamma/(\gamma - \Gamma_0)$ and $\gamma  = (Z_1 + Z_2)/(2Z_0 + Z_1 + Z_2)$ as a complex fitting parameter. The amplitude of $S_{11}$ and the location of its center of the resonance circle in the IQ plane will depend on the value of $\gamma_0$. \modifysecond{For a matched resonator, i.e. $Z_1 = Z_2 \rightarrow 0$, } $\gamma_0\rightarrow0$. $S_{11}$ will show a maximum at the resonance frequency. When \modifysecond{$Z_2 \rightarrow \infty$ and $Z_1 \rightarrow 0$, $\gamma_0\rightarrow 1/2$}, the magnitude of $S_{11}$ will be similar to $S_{21}$ and shows a minimum at the resonance. A more detailed mathematical procedure for Eq.~(\ref{eqn: S11 final}) can be found in Appendix A. 


\begin{figure*}
    \centering
    \includegraphics[width = \textwidth]{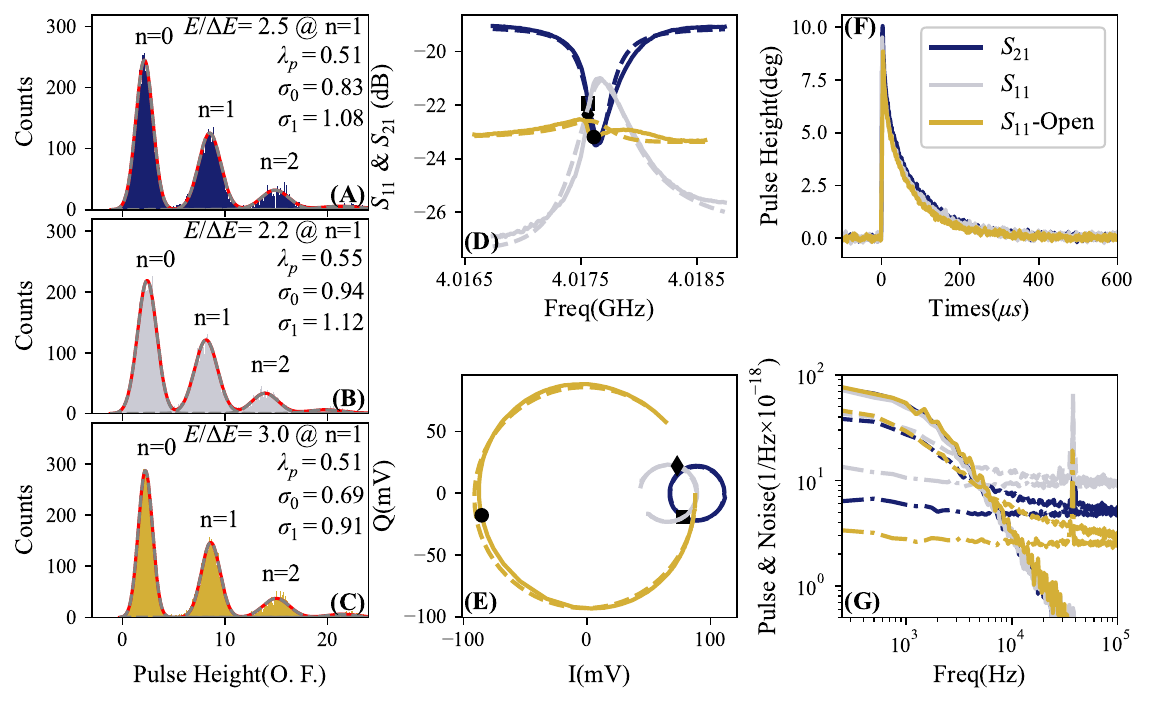}
    \caption{Comparison of the single-photon response of the MKID at 4.018~GHz from the transmission and reflection at $T_{bath} = 200~\text{mK}$ (A)-(C): Measured pulse statistics from $S_{21}$, $S_{11}$ and $S_{11}$-Open. $\lambda_p$ is the fitted mean value of the Poisson distribution. $\sigma_0$ and $\sigma_1$ are the standard deviations of the first and second peaks. (D)-(E): Comparison of the amplitude of the MKID and its trajectory on the IQ plane. The black dots indicate the readout frequency. \modify{The dashed lines are the fitted curves with Eq.~(\ref{eqn: S21}) and (\ref{eqn: S11 final}).} (F): The averaged single-photon response with hundreds of pulse events with $n=1$ as shown in (A)-(C). (G): Comparison between the noise spectrum and the pulse shown in (F), obtained by Fourier transformation. \modify{The dashed line ($--$) is the noise in phase, and the dash-dot line ($-\cdot$) is the noise in amplitude.}}
    \label{fig: Single-photon response comparison}
\end{figure*}

The detailed measurement setup for the MKIDs is illustrated in Fig. \ref{fig: Measurement Setup}. The MKIDs undergo characterization within a pre-cooled two-stage pulse tube Adiabatic Demagnetization Refrigerator (ADR). A 1.5mm-thick niobium cylinder, sheets of metglas 2714a surrounding the MKIDs, and a 2mm-thick cylinder of Cryophy are employed on the 4K stage to shield from magnetic interference. Each MKID is individually read out using a standard homodyne mixing scheme.

The input signal, generated by a signal generator, is attenuated by a programmable attenuator at room temperature and a 20~dB attenuator at 4~K. A circulator is positioned at the input of the MKIDs to capture the $S_{11}$ signal. Due to space limitations on the 100~mK stage, attenuators on the 1~K and 100~mK stages are temporarily removed to acquire the $S_{11}$ signal. \modifysecond{The thermal noise above 100~GHz is well-attenuated thanks to the thermal filters on the 100~mK stage. Thus, the background noise at the MKIDs should remain low\cite{Baselmans2012}, although the thermal noise at the resonance frequency band remains. It should also be noted that placing the circulator on the 100~mK stage will reduce the system noise, thus further improving $R$.}

The transmitted and reflected signals from MKIDs are initially amplified by an LNA  on the 4~K stage and further amplified by two room-temperature amplifiers. Subsequently, the transmitted and reflected signals are down-converted to DC by two IQ mixers and then sampled by an oscilloscope. The readout power at the detector's input is estimated to be approximately -100~dBm. 

Illumination of the MKIDs array is achieved through an optical fiber positioned 35~mm above the pixels. A 405~nm laser, modulated by a 250~Hz pulse from a pulse generator with a width of around 50~ns, is employed for illumination. The laser output power is adjusted outside the cryostat using a digital step attenuator. The pulse response of the MKID is sampled by an oscilloscope at a frequency of 5-10~MHz.

Two different configurations are employed for the measurement of the MKIDs. First, $S_{11}$ and $S_{21}$ are measured simultaneously. Second, the responses of the MKIDs terminated with an open load at the output port of the array are measured, as depicted in Fig. \ref{fig: Measurement Setup}, and the corresponding result is labeled as $S_{11}$-open.

Our array comprises 30 $\times$ 30 MKIDs. Each MKID consists of a 10~nm/10~nm/10~nm-thick TiN/Ti/TiN trilayer meander connected to a niobium capacitor. 
The meander size is $36 \times 36~\mu \text{m}^2$. The measured critical temperature $T_c$ of the trilayer film is around 1.75~K. A more detailed description of the design of the pixels can be found in our previous publication\cite{Hu2023}. 

As depicted in Fig.\ref{fig: Single-photon response comparison}, we first conducted a comparison of the measured single-photon response of the MKID obtained from $S_{11}$, $S_{21}$, and $S_{11}$-Open. In all cases, the MKID resonating at around 4.018~GHz was measured with a readout of approximately 2 dB below saturation at the frequency showing the maximum response on the IQ plane. Each measurement recorded around 10,000 pulse events and was processed by the standard wiener filtering\cite{Gao2012}. The resulting pulse height statistics are illustrated in Fig.\ref{fig: Single-photon response comparison}-(A)-(C), and these data were fitted using the convolution between the Gaussian and Poisson distributions, as\cite{Mezzena2019}.
\begin{align}\label{eqn: pulse statistics}
    p(\phi) = \sum_{n = 0}^{\infty} \frac{\lambda_p^n}{n!} e^{-(\phi-n\phi_1 - \phi_0)^2/(2\sigma_n^2)}
\end{align}
Here, $\lambda_p$ is the mean value of the Poisson distribution, $\phi_0$ is the mean value of the 0-th peak that is the background response from the chip, $\sigma_n$ is the variance of the n-th peak, and $\phi_1$ is the difference between the peaks, which is the response of a single photon.

\begin{figure*}
    \centering
    \includegraphics[width = \textwidth]{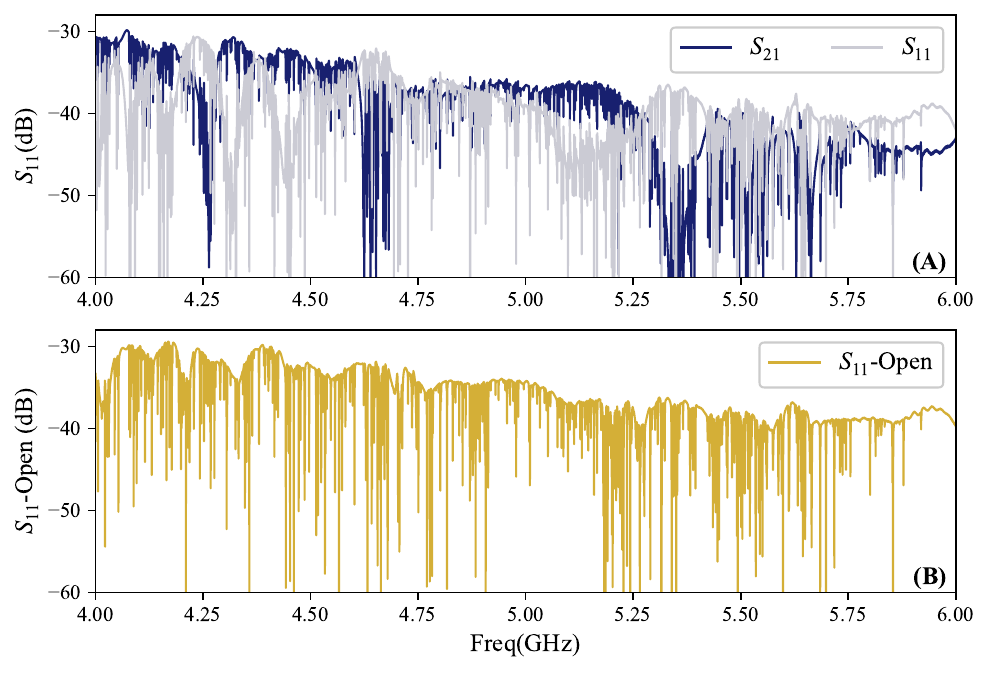}
    \caption{(A)-(B): Comparison of amplitudes of $S_{21}$, $S_{11}$ and $S_{11}$-Open respectively. There are two transmission zeros around 4.25~GHz and 5.35~GHz.}
    \label{fig: full bandwidth S11-s21-s11-open comparison}
\end{figure*}
\modify{Here, we consider the width of the 0-photon peak is dominated by the system noise $\sigma_s$, i.e., the two-level system (TLS) noise $\sigma_{TLS}$ from the MKID itself and the readout noise $\sigma_{r}$. 
\begin{align}
    \sigma_0^2 &= \sigma_{TLS}^2 + \sigma_{r}^2,
\end{align}
The width of the 1-photon peak is dominated by the readout noise and the noise related from the photon absorption $\sigma_{p}$, i.e., the hot phonon escape\cite{visser2021} and the current non-uniformity in the meander\cite{Zobrist2019} as 
\begin{align}
    \sigma_1^2 & = \sigma_0^2 + \sigma_{p}^2
\end{align}
}
\modify{Here, $\sigma_p$ is dominated by the phonon escape noise. Since the TLS is mainly in the phase response of MKID, $\sigma_r$ can be obtained by convolving the amplitude response in the 0-photon peak with the wiener filter.}
We list the contribution of each noise \modifysecond{of the resonator at 4.0175~GHz} in Table~\ref{table: Noise composition}. It can be seen that the $\sigma_{TLS}$ and $\sigma_p$ are consistent with each other. The difference in the $R$s of $S_{21}$, $S_{11}$ and $S_{11}$-open is due to the different levels of the readout noise. The $R$ of $S_{11}$-Open is dominated by the TLS and the phonon escape noise.

\begin{table}[]
    \centering
    \caption{\modifysecond{Fitted parameters and contribution of each noise to $R$ for the resonator at 4.0175~GHz.}}
    \begin{tabular}{ 
    >{\centering\arraybackslash}p{0.2\columnwidth} >{\centering\arraybackslash}p{0.09\columnwidth} >{\centering\arraybackslash}p{0.09\columnwidth} >{\centering\arraybackslash}p{0.2\columnwidth}  >{\centering\arraybackslash}p{0.1\columnwidth}  >{\centering\arraybackslash}p{0.1\columnwidth}  >{\centering\arraybackslash}p{0.1\columnwidth}}
    \hline\hline
        & $Q_c$& $Q_i$ &$\gamma_0$& $\sigma_{TLS}$  & $\sigma_{r}$ &  $\sigma_{p}$ \\ 
               & (k) & (k) &     &(O.~F.) &(O.~F.)&(O.~F.) \\ \hline
      $S_{21}$      & 32k & 21k & 1.0 &  0.61 &  0.56  &  0.61 \\ \hline
      $S_{11}$      & 25k & 17k & 0.06 + 1.0j &  0.60  &  0.72  &  0.69 \\ \hline
      $S_{11}$-Open & 16k & 19k & 0.27+0.35j & 0.58  &  0.38  & 0.59      \\ \hline
    \end{tabular}
    \label{table: Noise composition}
\end{table}

The increase in the readout noise in $S_{11}$ can be attributed to two possible reasons. 
First, the presence of the reflection chain especially the circulator can lead to an increase in readout noise. Second, the MKID is read out at the frequency exhibiting maximum response in the IQ plane, resulting in similar output powers at the output port of the MKID across all cases, as depicted in Fig.~\ref{fig: Single-photon response comparison}-(D)-(E), where the black dots represent the position of the readout signals. Consequently, the signal obtained from the reflection may not be significantly higher than the $S_{21}$'s. 
\modify{The dashed lines are the fitted results with Eq.~(\ref{eqn: S21}) and Eq.~(\ref{eqn: S11 final}) and the fitted parameters are listed in Table \ref{table: Noise composition}, which shows the internal quality factor $Q_i$ are in good agreement with each other. The initial values strongly influence the fitted parameters, yet their impact on estimating the pulse response is minimal.}

The reduction of the readout noise in $S_{11}$-open is due to the increase of the radius of the resonance circle $r$, which is shown in Fig.\ref{fig: Single-photon response comparison}-(E). as there is no significant difference in the pulse response shown in 
Fig.\ref{fig: Single-photon response comparison}-(F). The readout noise is related to $r$ as\cite{BARENDS2009Thesis} 
\begin{align}\label{eqn: S_DSB}
    S_{DSB} = \frac{k_B T_r }{r^2P_r},
\end{align}
where $k_B$ is the Boltzmann constant $T_r$ is the equivalent input noise temperature of the readout system. $P_r$ is the level of the readout power. Thus, the readout noise would be reduced when there is an increase in the radius of the resonance circle. This is also supported by the noise spectrum shown in Fig.~\ref{fig: Single-photon response comparison}-(G). 
Eq.~(\ref{eqn: S_DSB}) is qualitatively in agreement with the measurement by taking into account that \modify{the readout power of $S_{11}$-open is 2~dB lower than that of $S_{21}$}. For matched resonators (i.e, $Z_1 = 0$ and $Z_2 = 0$ for the case of $S_{11}$ and $S_{21}$, and $Z_1 = 0 $ and $Z_2 = \infty$ for $S_{11}$-Open), the radius of the resonance circle of $S_{11}$, $S_{21}$ and $S_{11}$-open can be expressed as 
\modify{
\begin{align}\label{eqn: radius of the resonance circle}
    r = \begin{cases}
        \frac{Q_i}{2(Q_i + Q_c)}\sqrt{4(Q_cx_a)^2 +1} & S_{11}~\&~S_{21} \\
        \frac{Q_i}{Q_i + Q_c^\prime}\sqrt{4(Q_c^\prime x_a)^2 +1} & S_{11}-\text{Open}
    \end{cases}
\end{align}
}
\modify{$Q_i$ is the internal quality factor of the resonator and $1/Q = 1/Q_c + 1/Q_i$. The coupling quality factor $Q_c^\prime$ of the $S_{11}$-Open  will change \sout{due} as 
\begin{align}\label{eqn: 14}
    Q_c = \frac{C}{\omega_rC_c^2}\Re{\frac{2Z_0 + Z_1 + Z_2}{(Z_0 + Z_1)(Z_0+Z_2)}}
\end{align}
where $C$ and $C_c$ are, respectively, the capacitors of the resonance and the coupling capacitor to the feedline. $\omega_r$ is the angular resonance frequency. $Q_c^\prime$ will reduce when $Z_0 + Z_2\rightarrow\infty$, which is agreement with the $Q_c$ fitted from the $S_{21}$ and $S_{11}$-open. In this case, for a matched resonator ($x_a = 0$), the radius of the resonance circle will be increased by more than a factor of 2.} A comprehensive mathematical derivation for determining the radius of the resonance circle can be found in Appendix B. 
 
\begin{figure}
    \centering
    \includegraphics[width = \columnwidth]{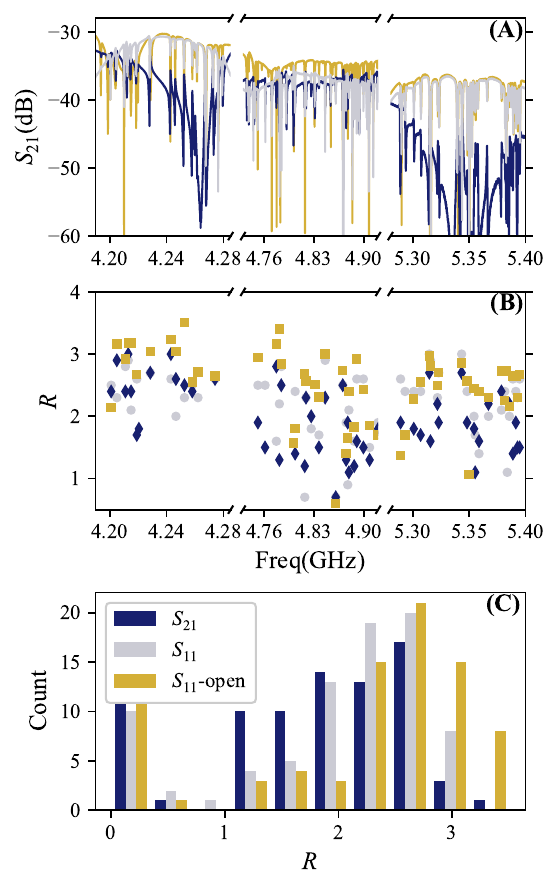}
    \caption{Comparison of the energy resolution between the $S_{11}$, $S_{21}$, and $S_{11}$-open over a bandwidth of around 100~MHz at 4.19-4.29~GHz, 4.75-4.95~GHz and 5.30-5.40~GHz. (A): The measured amplitude of the $S_{11}$, $S_{21}$ and $S_{11}$-open. (B): The measured energy resolution versus frequency. The input power is optimized for each pixel to maximize $R$. (C): Statistics of $R$ in the measured bands, around 10 pixels in the measured bands, do not show a single photon response.}
    \label{fig: energy resolution over bandwidth 100MHz}
\end{figure}
The readout scheme utilizing $S_{11}$-Open presents a notable advantage for systems featuring significant ripples in the baseline 
\modify{particularly due to the absence of bridges that balance the CPW ground\cite{Meeker2017}}.
In Fig.~\ref{fig: full bandwidth S11-s21-s11-open comparison}, we compare the amplitudes of $S_{11}$, $S_{21}$, and $S_{11}$-Open across the frequency range from 4.00~GHz to 6.00~GHz. Notably, where a minimum exists in $S_{21}$, there is a corresponding maximum in $S_{11}$, and vice versa. Additionally, as the frequency surpasses 5.50 GHz, the magnitude of $S_{11}$ begins to exceed that of $S_{21}$, elucidating the unexpected additional reduction in the magnitude of $S_{21}$ within our system. 
It can be seen that terminating the array with an open load remarkably diminishes the grounding effect to approximately $\pm3$ dB, a level comparable to state-of-the-art MKID arrays \cite{Baselmans2017, Shu2018}. 
\modifysecond{Note the amplitude of the $S_{11}$ and $S_{11}$-Open depends strongly on the $\gamma_0$ in Eq.~\ref{eqn: S11 final}, which is determined by the matching condition of the resonator to the circuit. When $\gamma_0$ is real positive, $S_{11}$ and $S_{11}$-Open will have similar amplitude with $S_{21}$. The amplitude of $S_{11}$-Open shown in Fig.\ref{fig: Single-photon response comparison} is a special case with the center of the resonance circle close to the origin of the IQ plane.}

Further investigations will be performed for larger MKIDs with multiple readout lines to check if the bridges that balance the ground can be eliminated.

We conducted a comparison of the energy resolution ($R$) across three frequency bands: 4.19-4.29 GHz, 4.75-4.95 GHz, and 5.30-5.40 GHz, where two of them exhibit a minimum in the amplitude, as depicted in Fig. \ref{fig: energy resolution over bandwidth 100MHz}-(A). The measured $R$s within the corresponding bands are illustrated in Fig. \ref{fig: energy resolution over bandwidth 100MHz}-(B). Despite the amplitude of $S_{21}$ being approximately 10 dB lower than that of $S_{11}$, the measured $R$ from $S_{21}$ remains comparable to $S_{11}$, while $R$ from $S_{11}$-Open demonstrates a significant improvement. This trend is consistent for the resonances observed in the 4.75-4.95 GHz and 5.30-5.40 GHz bands.

The statistical analysis of the $R$s depicted in Fig. \ref{fig: energy resolution over bandwidth 100MHz}-(B) is presented in Fig. \ref{fig: energy resolution over bandwidth 100MHz}-(C). Notably, the median $R$ is enhanced from 2.0 with $S_{21}$ to 2.6 with $S_{11}$-Open, and the maximum $R$ reaches approximately 3.5. 

\modify{
The impedance transformation of the readout line, denoted as ($Z_0$ + $Z_2$) in Eq.~(\ref{eqn: 14}), could potentially approach zero, leading to an increase in readout noise as $r$ decreases, as described in Eq.~(\ref{eqn: S_DSB}) and Eq.~(\ref{eqn: radius of the resonance circle}). To mitigate this issue, one effective approach is to position the resonators at multiples of half wavelength to the open load. Typically, an MKID array exhibits resonances between 4~GHz and 8~GHz, with the wavelength of an 8~GHz signal measuring around 16~mm on sapphire. By arranging the pixels within $\pm1~$mm around the half wavelength, approximately 10 pixels can be accommodated. However, achieving MKIDs in the optical band with resonance frequencies at desired positions presents challenges due to the small size of the pixels. Further investigations are warranted to optimize the performance of MKIDs with a reflection readout scheme.
Moreover, it's intriguing to note that despite our MKIDs not being designed to be terminated by an open load, we're observing an increase in the median $R$-value, suggesting the effectiveness of our method. Additionally, we believe that the proposed readout scheme remains applicable for MKID arrays in the millimeter bands, as indicated by the validity of Eq.~(\ref{eqn: S_DSB}) and Eq.~(\ref{eqn: radius of the resonance circle}). Even if there's a slight deviation in the resonance frequency from the design, it should still be manageable\cite{Shu2021_APL}.
}

In conclusion, we have implemented an MKID readout scheme based on reflection. Our observations indicate no significant improvement in the median $R$ when $S_{11}$ is measured together with $S_{21}$. Adopting a scheme to read out the single-end MKID array yields an enhancement in the median $R$ of the single-end MKID array by 20-30\%. This improvement is primarily attributed to the increase in the radius of the resonance circle. It is worth mentioning that this method can also be implemented in systems utilizing superconducting parametric amplifiers \cite{Zobrist2019, Zobrist2022}.


\section*{Availability of data}
The data supporting the findings of this study are available from the corresponding authors upon reasonable request.

\section*{AUTHOR DECLARATIONS
}
\section*{Conflict of Interest
}
The authors have no conflicts to disclose.

\begin{acknowledgments}

The authors would like to thank Shibo Shu from The Institute of High Energy Physics of the Chinese Academy of Sciences for the suggestion for the MKIDs measurement with reflection, as well as Florent Reix, Josiane Firminy, and Thibaut Vacelet from Paris Observatory for
assembly and mounting the devices. This work is supported
by the European Research Council (ERC) through Grant  835087 (SPIAKID) and the UnivEarthS Labex
program.
\end{acknowledgments}

\appendix

\renewcommand{\theequation}{S\arabic{equation}}

\section{Reflection of the Resonator with Mis-match}

The $Z_{in}$ of the system can be calculated as 
\begin{align}
    Z_{in} &= Z_1 + \frac{Z_3 (Z_2 + Z_0)}{Z_3 + (Z_2 + Z_0)} 
\end{align}
The $S_{11}$ can be obtained as 
\begin{align}
    S_{11} & = \frac{Z_{in}-Z_0}{ Z_{in}+Z_0} =\frac{Z_3(Z_1 + Z_2) + (Z_1-Z_0)(Z_0 + Z_2) }{Z_3(2Z_0 + Z_1 + Z_2) + (Z_0 + Z_1)(Z_0 + Z_2)}
\end{align}
with 
\begin{align}
    \hat{Z}_0 = \frac{2(Z_1 + Z_0)(Z_2 + Z_0)}{2Z_0 + Z_1 + Z_2}
\end{align}
Then, $S_{11}$ is \label{eqn: S11 transformed 1}
\begin{align}
    S_{11} = \frac{A Z_3 + \Gamma_0}{1 + 2Z_3/\hat{Z_0}}
\end{align}
with 
\begin{align}
    A &= \frac{Z_1 + Z_2}{(Z_0 + Z_1)(Z_0 + Z_2)} \\
    \Gamma_0 &= \frac{Z_1-Z_0}{Z_1+Z_0}
\end{align} 
The $S_{21}$ can be expressed as 
\begin{align}\label{eqn: S21}
    S_{21} = \frac{2Z_0}{2Z_0 + Z_1 + Z_2}(1 - \frac{1}{1 + 2Z_3/\hat{Z_0}})
\end{align}
with
\begin{align}
    \frac{1}{1 + 2Z_3/\hat{Z}_0} = \frac{Q/Q_c - 2jQx_a}{1+2jQx_g}
\end{align}
$S_{11}$ can be segmented into two parts as
\begin{align}\label{eqn: S11 two part}
    S_{11} = \gamma(1 -\frac{1}{1 + 2Z_3/\hat{Z}_0}) + \frac{\Gamma_0}{1 + 2Z_3/\hat{Z}_0}
\end{align}
with $\gamma = (Z_1 + Z_2)/(2Z_0 + Z_1 + Z_2)$.

By taking into account of Eq.~(\ref{eqn: S21}), Eq.~(\ref{eqn: S11 two part}) becomes 
\begin{align}
    S_{11} = \frac{Z_1 + Z_2}{2Z_0} S_{21} - \Gamma \frac{Q/Q_c - 2jQx_a}{1+2jQx_g}
\end{align}
or 
\begin{align}
    S_{11} &= \gamma - (\gamma - \Gamma_0)\frac{Q/Q_c - 2jQx_a}{1+2jQx_g}
\end{align}
Then, the $S_{11}$ can be fitted with 
\begin{align} \label{eqn: S11 final}
    S_{11} = (\gamma - \Gamma_0) ( \gamma_0 - \frac{Q/Q_c - 2jQx_a}{1+2jQx_g})   
\end{align}
with $\gamma_0 = \gamma/(\gamma - \Gamma_0)$ as a complex fitting parameter. 
For the single-end resonator, $Z_2\rightarrow\infty$ and $\gamma \rightarrow1$. If the resonator is perfectly matched, $Z_1 = 0$ and $\Gamma_0 = -1$. Thus, $S_{11}$ can be expressed as 
\begin{align}
    S_{11} = 1-2\frac{Q/Q_c - 2jQx_a}{1+2jQx_g}
\end{align}
which is similar with the expression of $S_{11}$ found in previous publications. 

\section{The Coupling Quality Factor and Resonance circle}
The coupling quality factor can be calculated as follows. The energy stored in the capacitor and inductor are the same at the resonance frequency, which is
\begin{align}
    E = 2\times \frac{1}{2} CV^2
\end{align}
where $V$ is the voltage of the capacitor. The energy dissipated outside the resonator is 

\begin{align}
    P = |I|^2\Re{R}\approx \omega^2C_c^2V^2\Re{\frac{(Z_0 + Z_1)(Z_0 + Z_2)}{2Z_0 + Z_1 + Z_2}}
\end{align}

$Q_c$ can then be calculated by definition as 
\begin{align}
    Q_c = \omega \frac{E}{P} = \frac{C}{\omega C_c^2} \Re{\frac{2Z_0 + Z_1 + Z_2}{(Z_0 + Z_1)(Z_0 + Z_2)}}.
\end{align}

We compare the radius of the resonance circle between the $S_{11}$, $S_{21}$ and $S_{11}$-Open. The term in the bracket of  Eq.~(\ref{eqn: S11 final}) 
\begin{align}
    S_{11}^b = \gamma_0 - \frac{Q/Q_c - 2jQx_a}{1+2jQx_g}
\end{align}
is a circle on the IQ plane. The three points on the resonance circle can be determined as 
\begin{align}
    S_{11}^b = \begin{cases}
        \gamma_0 & x\rightarrow\infty \\ 
        \gamma_0 - Q/Q_c -2jQ x_a & x= 0 \\ 
        \gamma_0 - \frac{Q/Qc - 2jQx_a}{1 + jQ} & x = 1/2 \\
    \end{cases}
\end{align}
Here we assume $\gamma_0 = \gamma_r + j\gamma_i$, then the center of and the radius of the resonance circle is 
\begin{align}
    [x_0^b,~y_0^b] & = [\gamma_r - \frac{Q}{2Q_c}, \gamma_i + 2Qx_a] \\
    R_b & = \frac{Q}{2Q_{c}}\sqrt{4(Q_cx_a)^2 + 1}
\end{align}
Taking into account of the coefficients, the radius of the resonator circle of the $S_{11}$ is 
\begin{align}
    R_r = |\gamma_{21} - \Gamma_0|\frac{Q}{2Q_{c}}\sqrt{4(Q_cx_a)^2 + 1}
\end{align}
If the resonator is perfectly matched, $\gamma_{21} = 0$. $\Gamma_0 = -1$. In the case that the $Z_2\rightarrow\infty$, which is the case that the MKIDs is terminated with an open load.  $\gamma_{21} = 1$ In the case 
\begin{align}
    R_r = \begin{cases}
        \frac{Q_i}{2(Q_i + Q_c)}\sqrt{4(Q_cx_a)^2 + 1} &   Z_1 = 0 ~~\&~~ Z_2 = 0 \\
        \frac{Q_i}{Q_i + Q_c^\prime}\sqrt{4(Q_c^\prime x_a)^2 + 1} &   Z_1 = 0 ~~\&~~ Z_2 = \infty 
    \end{cases}
\end{align}
with $1/Q = 1/Q_i + 1/Q_c$. Thus, there is an increase of a factor of 2 in the resonator terminated with an open load. Similarly, the radius of $S_{21}$  
\begin{align}
    R_t = |\frac{2Z_0}{2Z_0 + Z_1 + Z_2}|\frac{Q}{2Q_{c}}\sqrt{4(Q_cx_a)^2 + 1}
\end{align}
In the situation that $Z_1 = Z_2 = 0$, 
\begin{align}
    R_t = \frac{Q_i}{2(Q_i + Q_c)}\sqrt{4(Q_cx_a)^2 + 1}
\end{align}

\bibliography{text.bib}

\end{document}